# An eccentric disk model for the nucleus of M31

SCOTT TREMAINE

Canadian Institute for Theoretical Astrophysics, University of Toronto,
60 St. George St., Toronto M5S 1A7, Canada[1]
and Institute of Astronomy, Madingley Road, Cambridge CB3 0HA, UK
Electronic mail: tremaine@cita.utoronto.ca

## ABSTRACT

The nucleus of M31 may be a thick eccentric disk, composed of stars traveling on nearly Keplerian orbits around a black hole or other dark compact object. This hypothesis reproduces most of the features seen in HST photometry of the center of M31; in particular the bright off-center source P1 is the apoapsis of the disk. An eccentric disk can also explain the rotation curve and asymmetric dispersion profile revealed by ground-based observations. The central object must be smaller than $\sim 1\,\mathrm{pc}$ so that the potential felt by the disk is nearly Keplerian. The disk eccentricity may be excited by a secular instability driven by dynamical friction from the bulge.

[1]Permanent address

# 1 Introduction

The nucleus of M31 may be the nearest dead quasar, that is, it may contain a black hole (hereafter BH) of sufficient mass ($\sim 10^8\,\mathrm{M}_\odot$) to power an active galactic nucleus when offered an adequate fuel supply. The nucleus was first resolved in high-resolution (FWHM=$0\farcs2$) photographs by the Stratoscope II balloon-borne telescope (Light, Danielson & Schwarzschild 1974), which also revealed that the peak brightness of the nucleus was offset relative to its outer parts.

Hubble Space Telescope (HST) images with FWHM=$0\farcs08$ show that the nucleus contains two separate components, separated by $0\farcs49$ (Lauer et al. 1993, hereafter L93; see also King, Stanford & Crane 1995). The component with the lower surface brightness, P2, coincides with the center of the outer isophotes of the nucleus and the bulge photocenter to within $\sim 0\farcs05$, and exhibits a shallow central cusp. The brighter component, P1, is well-resolved and coincides with the nuclear core measured by Stratoscope. No strong color gradients are visible between the near-infrared and the far ultraviolet, except for enhanced emission in the far ultraviolet close to P2 (Mould et al. 1989; King et al. 1995).

Ground-based spectroscopy (Dressler & Richstone 1988; Kormendy 1988; Bacon et al. 1994; van der Marel et al. 1994) shows that the rotation speed rises from near zero outside $4''$ to a maximum of $\simeq 120\,\mathrm{km\,s^{-1}}$ at $1''$ from the center. The location and amplitude of the rotation maximum are determined by the spatial resolution, so the actual maximum is almost certainly larger. The center of symmetry of the rotation curve coincides with P2 to within $0\farcs2$; the curve is symmetric so there is no evidence that the brighter peak P1 perturbs the rotation curve. The velocity dispersion rises from $\simeq 150\,\mathrm{km\,s^{-1}}$ outside $2''$ to a maximum of $220\,\mathrm{km\,s^{-1}}$; once again the actual maximum is certainly larger. The dispersion peak is displaced from P2 to the anti-P1 side and the dispersion profile is not symmetric.

Kinematic models—which usually neglect these asymmetries—require that the mass in the central arcsecond is dominated by a BH or other compact non-luminous mass concentration, with mass $M_\bullet \simeq 10^{7.5}\,\mathrm{M}_\odot$ (Kormendy 1988; Dressler & Richstone 1988; Bacon et al. 1994).

Explanations of the asymmetries in the M31 nucleus are discussed by L93 and Bacon et al. (1994). Partial obscuration by dust is implausible because there are no color gradients or re-radiated far-infrared emission from the dust; also, if P2 is obscured by an asymmetric dust distribution, why should it be exactly at the center of the bulge? Bacon et al. (1994) suggested that P1 is off-center because of a dynamical oscillation, but no specific model was proposed so their suggestion cannot be assessed.

A natural hypothesis is that P1 is a separate stellar system orbiting the nucleus. However, assuming P1 is close to the nucleus—the probability of a chance superposition of a foreground or background object is negligible—its orbit rapidly decays by dynamical friction from the bulge. For the best-fit bulge model described below, the frictional decay time[2] for an object of mass $m$ is $\sim 6\times 10^7\,\mathrm{yr}(10^6\,\mathrm{M}_\odot/m)$, for radii between $0\farcs5$ and $3''$ from the center (for comparison the orbital time at $1''$ is $8\times 10^4$ yr). Thus the present configuration is short-lived. Most attempts to evade this conclusion are implausible: depletion of the stars that contribute to the friction does not occur because tidal forces in the triaxial bulge (Stark 1977) continually bring fresh stars close to the center; compensating the friction by energy transfer from a disk is only a temporary fix, because the reservoir of energy in the disk is too small. Finally, tidal forces from the central BH should disrupt P1 unless its mass-to-light ratio is substantially higher than a normal stellar population.

---

[2] The object is assumed to be on a circular orbit and the decay time is $L/\dot L$ where $L$ is the angular momentum. The velocity distribution of bulge stars is assumed to be Maxwellian and the drag is computed using the usual Chandrasekhar formula with Coulomb logarithm $\ln\Lambda = \ln(3.0)$.

Some of these problems are avoided if P1 is a massive black hole surrounded by bound stars. Lacey & Ostriker (1985) have argued that the dark matter in galactic halos may be composed of BHs of mass $\sim 10^{6.5}\,\mathrm{M}_\odot$ (but see Moore 1993 and Carr 1994 for counter-arguments). Black holes of this mass will spiral to the center by dynamical friction from as far out as $\sim 1\,\mathrm{kpc}$, with one arriving every $10^8$ yr or so. The fate of these holes is controversial: (a) they may all coalesce into a dense cluster or even a single BH (Hut & Rees 1992); (b) gravitational interactions between a binary BH and a single BH may eject one or both systems, so that the number of BHs at the center is normally 0, 1, or 2 but no higher (Xu & Ostriker 1994). Thus the short lifetime of P1 is not a problem, because fresh black holes arrive at a steady rate; in fact, Xu and Ostriker show that a binary BH with separation similar to that of P1 and P2 is present about 10% of the time. However, it remains uncertain whether this model can naturally explain the coincidence of P2 with the center of the bulge and the symmetry of the rotation curve; also, acquiring the "cloak" of stars around the BH at P1 is difficult, since the phase-space density of stars around P1 is higher than anywhere else in the bulge or nucleus.

The purpose of this paper is to present a model for the central region of M31 that explains the observed asymmetries.

## 2 A model for the nucleus of M31

The model assumes that a single massive BH is located at the center of the bulge, which may be taken to be the center of P2 (apart from a small correction described below).

The nucleus is assumed to be an eccentric stellar disk, composed of stars traveling on Keplerian orbits around the BH at P2. The disk is eccentric because the apsides of the orbits are approximately aligned. Kormendy (1988) has already argued that the nucleus may be a disk; the only novelty here is that the disk is eccentric.

The goal of this section is to demonstrate that an eccentric disk can explain the main features of the M31 nucleus; the model is not designed to fit the observations to the highest possible accuracy. The distance to M31 is assumed to be 770 kpc (Freedman & Madore 1990) so that $1'' = 3.73\,\mathrm{pc}$, and the absorption correction is taken to be $A_V = 0\overset{\mathrm{m}}{.}24$. The mass of the BH is $M_\bullet = 7.5 \times 10^7\,\mathrm{M}_\odot$ and the mass-to-light ratio of the bulge is $M/L = 4\,\mathrm{M}_\odot/\mathrm{L}_\odot$ after correcting for absorption, values chosen to yield approximate agreement with the observed dispersion profile (see below). The nuclear stars are assumed to have the same mass-to-light ratio as the bulge. The mass-to-light ratio of the bulge is close to the estimate $5.6\,\mathrm{M}_\odot/\mathrm{L}_\odot$ given by Kent (1989, after correcting for absorption and converting from $r$ to $V$).

The bulge is assumed to be spherical (its actual ellipticity $\epsilon \simeq 0.15$–$0.20$; Kent 1989) and its surface brightness is described by the fitting formula

$$I(r) = 2^{(\beta-\gamma)/\alpha} I_b \left(\frac{r}{r_b}\right)^{-\gamma} \left[1 + \left(\frac{r}{r_b}\right)^\alpha\right]^{-(\beta-\gamma)/\alpha}, \qquad (1)$$

which reproduces the surface brightness distribution of a wide variety of elliptical galaxies and bulges (Lauer et al. 1995). The formula has five parameters: the three exponents $\alpha$, $\beta$, $\gamma$, the break radius $r_b$, and the surface brightness at the break radius, $I_b$. Rotation of the bulge is neglected. The velocity dispersion tensor in the bulge is assumed to be isotropic, and is determined by solving the Jeans equations in the combined potential field of the central BH and the bulge stars.

The nucleus (or "nuclear disk") is an eccentric disk lying in the plane defined by the disk of M31 (inclination 77°, position angle of the line of nodes 38°). The disk is represented by assuming that its stars lie on a small number of nested Keplerian orbits ("ringlets") with semi-major axes $a_n$, eccentricities $e_n$, and total luminosities $L_n$, $n = 1, \ldots, N$. The focus of each ringlet is at P2 and the apoapsis direction is assumed to coincide with the P2–P1 axis (position angle 43° according to L93, which implies that the azimuthal angle in the disk plane between the line of apsides and the line of nodes is 21°). To smooth the model and to account crudely for the disk thickness, the sky brightness distribution of each ringlet is convolved with a Gaussian point-spread function with standard deviation $\sigma_n$; thus $\sigma_n/a_n$ is an approximate measure of the eccentricity and inclination spread in the disk. The velocity field is determined from the Keplerian velocities of the ringlets, with no correction for asymmetric drift.

The model neglects the gravitational influence of the nuclear disk on the bulge, since its mass is only 16% of the BH mass in the best-fit model[3]; the model also neglects the influence of the bulge on the nuclear disk, since its mass interior to $1''$ is only 2% of the BH mass.

The parameters of the model are determined by fitting two data sets:

(a) The surface brightness measured by HST in a strip of width $0\rlap{.}''22$ along the P2–P1 axis, extending to $\pm 4''$ from P2 (see Figure 6 of L93).

(b) The surface brightness on the major axis of the bulge between $4''$ and $100''$ (Kent 1989). Kent's $r$-band magnitudes are increased by $0\rlap{.}^m 36$ so that they match the HST $V$-band magnitudes in the region of overlap.

The fitting procedure minimizes the mean square deviation in magnitude, giving equal weight to the two data sets. The model shown in this paper (Figure 1) uses only 3 ringlets, but already fits the data reasonably well (the rms residual in either data set is less than $0\rlap{.}^m 04$); models with more ringlets provided only modest improvements to the fit and are increasingly ill-conditioned. The model does not reproduce precisely the shape of the profile in the region between P2 at $x = 0$ and P1 near $x = -0\rlap{.}''6$; however, these discrepancies could probably be removed in more careful models and are nowhere greater than $0\rlap{.}^m 1$.

---

[3] The principal consequence of this approximation is that our estimate of the BH mass is probably too large by $\sim 10\%$.

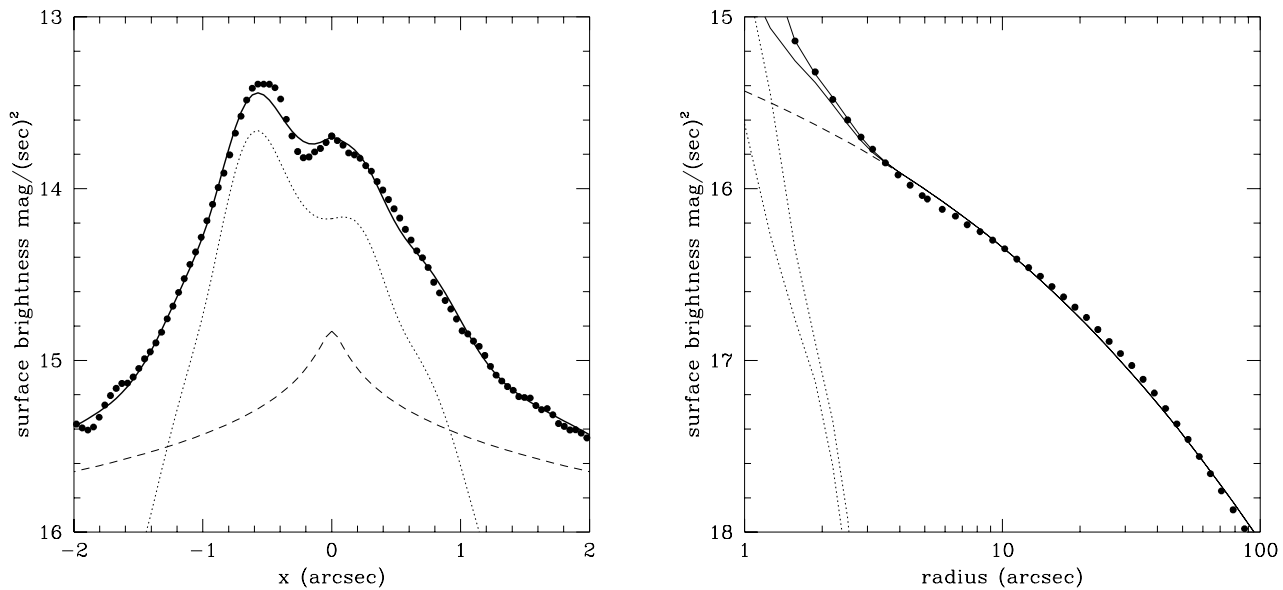

**1.** (a) V-band surface brightness in a slice $0''\!.22$ wide along the P2–P1 axis. P2 is at $x = 0$ and P1 is at $x = -0''\!.5$. The filled circles show the HST data (Figure 6 of L93), and the solid line shows the best-fit model with 3 ringlets. The dashed and dotted lines show the contributions from the bulge and the nuclear disk. (b) V-band surface brightness along the major axis of the bulge. The filled circles show $r$-band photometry from Kent (1989), adjusted by $0\overset{m}{.}36$ downward to agree with L93 in the region of overlap. The two solid lines show the best-fit model along both sides of the P2–P1 axis; the dashed and dotted lines show the contributions from the bulge and the nuclear disk.

The fit shown in Figure 1 corresponds to the parameters (luminosities are corrected for absorption but apparent magnitudes are not),

$$\alpha = 0.58, \quad \beta = 1.37, \quad \gamma = 0.15, \quad I_b = 17\overset{m}{.}42 \ (\sec)^{-2}, \quad r_b = 49''\!.1,$$

$$a_1 = 0''\!.49, \quad e_1 = 0.44, \quad L_1 = 5.8 \times 10^5 \ L_\odot = 15\overset{m}{.}09, \quad \sigma_1 = 0''\!.16,$$

$$a_2 = 0''\!.87, \quad e_2 = 0.15, \quad L_2 = 9.5 \times 10^5 \ L_\odot = 14\overset{m}{.}56, \quad \sigma_2 = 0''\!.29, \qquad (2)$$

$$a_3 = 1''\!.53, \quad e_3 = 0.04, \quad L_3 = 1.43 \times 10^6 \ L_\odot = 14\overset{m}{.}12, \quad \sigma_3 = 0''\!.78.$$

The total luminosity of the disk is $3.0 \times 10^6 \ L_\odot = 13\overset{m}{.}3$, although the peak at P1 is mostly due to ringlet 1 which is fainter. Many of these parameters are rather uncertain as the inversion is ill-conditioned.

Figure 2 shows the surface brightness distribution predicted by the model, which can be compared to Figure 2 of L93. The model successfully reproduces the main features of the HST image, even though the fit was based on data from only the P2–P1 axis. There are minor discrepancies—for example, the ellipticity of the isophotes surrounding P1 and P2 is higher in the model than in the data—but these should be easy to remove in more careful models.

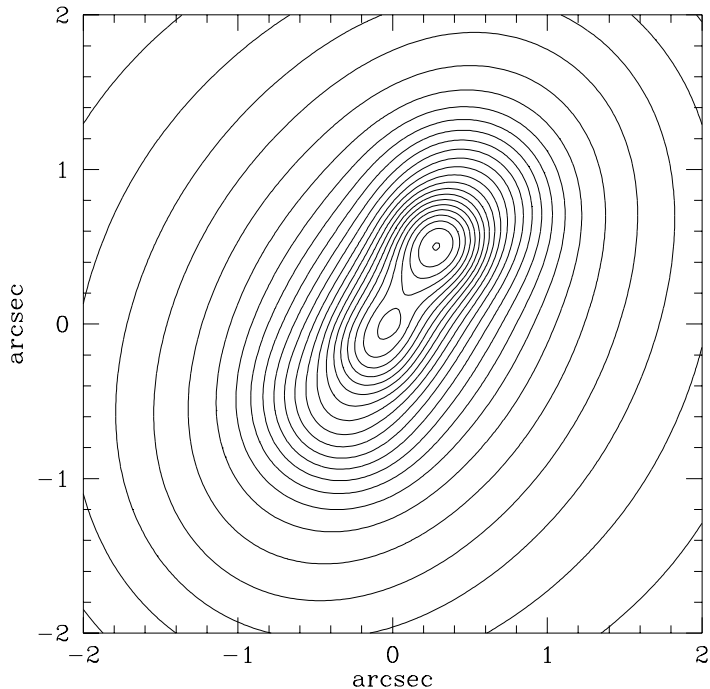

2. Contour map of the surface brightness of the best-fit model. The contour interval is $0\overset{m}{.}1$. The map can be compared to Figure 2 of L93. The vertical axis points 70° counterclockwise from north.

An eccentric disk may explain other features observed by HST (L93; King et al. 1995): (a) the region between P1 and P2 is darker than would be expected from the superposition of two elliptical components (there may be a vacant region inside the nuclear disk); (b) the P1 isophotes are twisted (isophote twists are a natural consequence of an eccentric disk whose eccentricity or thickness varies with radius); (c) at the highest surface brightness levels, both P1 and P2 appear elongated along the P2–P1 axis, a natural feature of a disk viewed at a high inclination angle; (d) P2 exhibits a weak cusp and enhanced ultraviolet emission at its center, as might be expected close to a BH, while P1 does not, as expected for a disk.

The center of the bulge should coincide with the center of mass of the BH and the disk rather than precisely with the BH (i.e. P2). Thus in the best-fit model the center of the bulge should be displaced from P2 towards P1 by $0\overset{''}{.}02$. King et al. (1995) estimate that the center of the bulge is displaced from P2 towards P1 by $0\overset{''}{.}05$, while L93 estimate that the displacement is between 0 and $0\overset{''}{.}05$.

Figure 3 shows the mean velocity and velocity dispersion as a function of distance from P2[4]. The left-hand panels show data from Kormendy (1988) along position angles PA= 55° (filled squares) and 38° (open squares), along with predictions obtained by convolving the

---

[4] The comparison of theoretical dispersion profiles with measured dispersions can be misleading, especially when the light comes from stars with a range of velocity dispersions (the cool nuclear disk, the hot bulge, and the even hotter cusp of bulge stars near the BH). This complication is neglected in the present paper (see Kormendy 1988 and van der Marel 1994 for discussion).

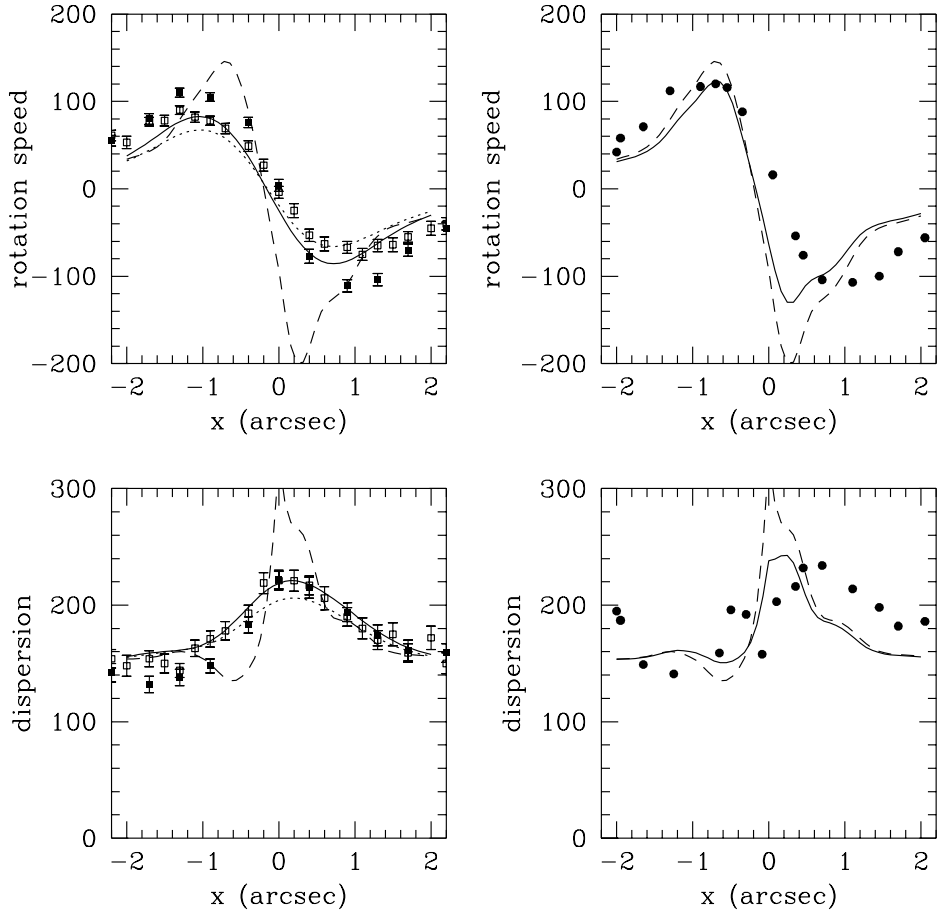

3. Rotation velocity and velocity dispersion as a function of distance from P2 in the best-fit model. The profiles have been "observed" with a Gaussian point-spread function of standard deviation $\sigma_\star$, through a slit of width $s$ at position angle PA. The curves in the left-hand panels have parameters that approximate Kormendy's (1988) observations: the solid curves have $\sigma_\star = 0\farcs52$, $s = 0\farcs5$, PA= $55°$, and the dotted curves have $\sigma_\star = 0\farcs65$, $s = 1\farcs0$, PA= $38°$. The corresponding observations are shown as filled and open squares, respectively. Note that positive $r$ in Kormendy (1988) corresponds to negative $x$, i.e. P1 is at $x \simeq -0\farcs5$. The profiles in the right-hand panels have $\sigma_\star = 0\farcs37$, $s = 0\farcs5$, PA= $54°$, corresponding to Bacon et al.'s (1994) observations which are shown as filled circles. The dashed lines show the rotation and dispersion curves that would be observed with perfect resolution.

best-fit model with the appropriate point-spread functions and slit widths. The profiles that would be seen at perfect resolution are shown as dashed lines. The right-hand panels show rotation and dispersion data from Bacon et al. (1994).

First consider the comparison to Kormendy's data. The model correctly predicts the approximate shape of the rotation profile, and the location of the zero-rotation point close to P2. The maximum rotation speed is too low by ∼20%, suggesting that the model underestimates the prominence of the nuclear disk. The model correctly predicts the shape of the dispersion profile—although the bulge mass-to-light ratio and the mass of the central BH have been chosen to optimize this fit. There are slight asymmetries in the model rotation curve: the maximum of the rotation curve on the P1 side ($x < 0$) is at a larger distance than on the anti-P1 side ($-1''$ compared to $+0\rlap.{''}8$), and the zero-rotation point is displaced $0\rlap.{''}1$ towards P1 from P2. There is no clear asymmetric signal in the data, although asymmetries would be masked by Kormendy's decision to define the origin as the zero-rotation point of the rotation curve. The dispersion profile of the model is asymmetric, with dispersions on the anti-P1 side ($x > 0$) being larger than dispersions on the P1 side; this agrees with the sense of the asymmetry in the data although the asymmetry in the model is smaller.

The right-hand panels compare the model to data from Bacon et al. (1994). The model correctly predicts the amplitude of the rotation curve; the principal discrepancies are that (a) the linear part of the model rotation curve is too steep; (b) the zero-rotation point of the model curve is displaced from P2 towards P1 by about $0\rlap.{''}2$, while the analogous point in the data is displaced away from P1 by $0\rlap.{''}1$–$0\rlap.{''}2$. The model dispersion profile correctly predicts that the dispersion is enhanced on the anti-P1 side although the width of the dispersion peak is too small; as in the case of the rotation curve, the agreement would be improved if the model curve were shifted away from P1. These discrepancies are probably minor given the spatial resolution of the data (sampling of $0\rlap.{''}39$ and seeing FWHM of $0\rlap.{''}87$).

All of these kinematic comparisons are with a single model that was designed mainly to match the photometry along the P2–P1 axis. No kinematic information was used in the fit except that the mass of the central BH and the bulge mass-to-light ratio were adjusted to match the dispersion curve. Simultaneous fits of eccentric disk models to both the photometry and the kinematics have not been explored.

## 3 Discussion

All eccentric disk models that match the sky brightness distribution should share several features:

(a) If the disk lies in the same plane as the M31 disk (inclination $77°$) then it cannot be too thin, or else the isophotes surrounding P1 would be strongly flattened towards the P2–P1 axis. To fit the observations, the rms disk thickness $h$ must exceed roughly $0.3r$ where $r$ is the radius. In fact a thick disk is also required by dynamical arguments. The thickness grows on the two-body relaxation time, which is given by

$$t_{\rm relax} \approx \frac{P}{\ln\Lambda}\left(\frac{h}{r}\right)^4 \frac{\Delta r}{r} \frac{M_\bullet^2}{m M_d}, \qquad (3)$$

where $P = 2\pi(r^3/GM_\bullet)^{1/2}$ is the orbital period, $\Delta r$ is the radial width of the disk, $m$ is the typical stellar mass, $M_d$ is the disk mass, and $\ln\Lambda$ is the Coulomb logarithm.

Inserting approximate values for the M31 disk,

$$
\begin{aligned}
t_{\rm relax} = 10^{12.3}\,{\rm yr} &\left(\frac{h}{r}\right)^4 \left(\frac{10}{\ln\Lambda}\right)\left(\frac{r}{2.5\,{\rm pc}}\right)^{3/2} \\
&\times \left(\frac{\Delta r}{0.5r}\right)\left(\frac{M_\bullet}{10^{7.8}\,{\rm M}_\odot}\right)^{3/2}\left(\frac{1\times 10^7\,{\rm M}_\odot}{M_d}\right)\left(\frac{0.5\,{\rm M}_\odot}{m}\right).
\end{aligned}
\qquad (4)
$$

Thus the disk thickness $h$ must exceed $0.3r$ in order that the relaxation time is longer than the Hubble time.

(b) The surface brightness of the disk may be written $I(r,\phi)$ where $r$ is the radius in the disk and $\phi$ is the azimuthal angle. Suppose that the disk is cold, the streamline with semi-major axis $a$ has eccentricity $e(a)$ and periapsis angle $\varpi(a)$, the luminosity in the interval $[a, a+da]$ is $\ell(a)da$, and the azimuthal origin is chosen so that $\varpi = 0$ or $\pi$. Then $e$ and $\varpi$ may be replaced by the single variable $k(a) \equiv e(a)\cos\varpi(a)$ ($-1 \le k \le 1$), and

$$
\begin{aligned}
I(r,0) &= \frac{\ell(a)}{2\pi(1-k^2)^{1/2}} \frac{1-k}{1-k-adk/da}, \\
I(r,\pi) &= \frac{\ell(a)}{2\pi(1-k^2)^{1/2}} \frac{1+k}{1+k+adk/da},
\end{aligned}
\qquad (5)
$$

where $a$ is determined implicitly from $r(a,\phi=0) = a[1-k(a)]$ and $r(a,\phi=\pi) = a[1+k(a)]$. The condition that streamlines do not intersect is $|k+adk/da| < 1$. The surface brightnesses at a given semi-major axis on opposite sides are in the ratio

$$
\frac{I[r(a,\pi),\pi]}{I[r(a,0),0]} = \frac{1 - \dfrac{adk/da}{1-k}}{1 + \dfrac{adk/da}{1+k}}. \qquad (6)
$$

Now let $\phi = 0$ correspond to the direction from P2 to P1. Since the surface brightness at P1 is higher than anywhere on the anti-P1 side, the right side of equation (6) must be less than unity, which requires $dk/da > 0$. Thus, for example, if P1 coincides with the apoapsides of the disk streamlines, then $\varpi = \pi$, $k < 0$ and the eccentricity gradient $de/da$ must be negative in the brightest parts of the disk (consistent with the monotonic decrease of $e$ with increasing $a$ in the ringlets of equation 2).

(c) The density distribution along a Keplerian orbit can be expanded as a Fourier series in azimuth. If the eccentricity is small, the dominant azimuthal wavenumber is $m = 1$. The asymmetry associated with P1 has substantial components at azimuthal wavenumbers greater than 1; therefore the eccentricity of the disk in the region contributing to P1 must be substantial, at least 0.3 and probably greater.

A mechanism is needed to maintain apsidal alignment against differential (retrograde) precession caused by the bulge (in the best-fit model, differential precession between near-circular orbits at $0\rlap{.}''5$ and $1''$ is $\sim 1$ radian/Myr). One possibility is that the apsides are trapped at a preferred orientation by the triaxial potential from the bulge (Stark 1977). However, for modest levels of triaxiality only highly eccentric orbits are trapped; also, if the disk is composed of stars on trapped orbits with periapsis angle $\varpi$, then we must still explain why the equivalent trapped orbits with periapsis angle $\varpi + \pi$ are not also populated.

A more promising possibility is that the alignment is maintained by the self-gravity of the disk. The eccentric distortion can be regarded as a discrete nonlinear eigenmode of a self-gravitating disk orbiting in the external field of the BH and the bulge. The eigenmode has a pattern speed $\Omega_p$, the common apsidal precession rate, which is much less than the typical Keplerian angular speed $\Omega = (GM_\bullet/r^3)^{1/2}$ because the disk mass is much less than the BH mass. The properties of such disks have not been investigated much[5], and we would like to know under what circumstances they exist, and what are their shapes and pattern speeds.

A related problem is why the disk should be eccentric. Gurzadyan & Ozernoy (1979) and Syer & Clarke (1992) have suggested that eccentric accretion disks form through tidal disruption of an object passing by the BH (most likely a giant molecular cloud). Another possibility is that the eccentricity arises from an unstable $m = 1$ normal mode of the axisymmetric disk. A third mechanism, perhaps the most promising, is the influence of dynamical friction from the bulge on a nearly circular disk. Suppose for the moment that the bulge does not rotate and its velocity dispersion tensor is isotropic, so that dynamical friction removes energy from the disk, $\dot{E} < 0$ (Tremaine & Weinberg 1984). Dynamical friction also removes angular momentum, at a rate $\dot{J} = \dot{E}/\Omega_p$. Thus if $0 < \Omega_p \ll \Omega$, angular momentum is removed much faster than energy so the disk orbits become more eccentric: in other words there is a secular instability that causes the eccentricity of the disk to grow. On the other hand, if $\Omega_p < 0$ the friction adds angular momentum so the disk becomes less eccentric. Disks can have positive or negative pattern speeds: the bulge potential causes retrograde precession but the self-gravity of the disk causes prograde precession. Since the mass of the disk is larger than the mass of the bulge inside the disk, most or all eigenmodes have $\Omega_p > 0$ so their amplitudes grow. If the bulge rotates with mean angular speed $\Omega_b$, then only eigenmodes with $\Omega_p > \Omega_b$ will grow. An unsolved problem is what limits the growth of the secular instability.

The presence of an eccentric disk is difficult to reconcile with the arrival of $10^{6.5}\,M_\odot$ BHs every $\sim 10^8$ yr, as required if the halo is composed of such objects (Lacey & Ostriker 1985; Xu & Ostriker 1994), since each inspiraling BH would destroy the disk.

The model described here implies that the size of the dark object at the center of M31 should be $\lesssim 0''\!.25 \simeq 1\,\mathrm{pc}$ (the periapsis distance of ringlet 1) so that the force field seen by the eccentric disk is nearly Keplerian—otherwise the relatively weak self-gravity of the disk could not maintain apsidal alignment. This upper limit is more stringent than limits based on variation of mass-to-light ratio with radius (Richstone, Bower & Dressler 1990) and close to the lower limit for the size of a stellar cluster, which must have half-mass radius $\gtrsim 0''\!.1$ (Goodman & Lee 1989).

---

[5] Except for the case of narrow cold disks with sharp edges (Goldreich & Tremaine 1979).

Higher resolution spectroscopic observations, improved photometry from HST, and measurements of the shape of the line-of-sight velocity distribution can all provide sensitive probes of eccentric disk models in the near future. Equally important tasks are to develop a better theoretical understanding of the dynamics of collisionless eccentric disks, and self-consistent models of eccentric disks that predict the line-of-sight velocity distribution.

I thank my HST collaborators Ed Ajhar, Yong-Ik Byun, Alan Dressler, Sandra Faber, Karl Gebhardt, Carl Grillmair, John Kormendy, Tod Lauer, and Doug Richstone for many conversations and insights about galactic cores and nuclei. I also thank Roland Bacon, Gerry Quinlan, Jerry Sellwood, Jerry Ostriker, and Dave Syer for discussions about the nucleus of M31, and Jerry Ostriker for introducing me to this interesting system. This work was carried out during a visit to the Institute of Astronomy, Cambridge, which was made possible by assistance from the Killam Program of the Canada Council and the Raymond and Beverly Sackler Foundation. The research was also supported by grants from NSERC.

# References

Bacon, R., Emsellem, E., Monnet, G., & Nieto, J. L. 1994 A&A 281, 691

Carr, B. 1994, ARAA 32, 531

Dressler, A., & Richstone, D. O. 1988 ApJ 324, 701

Freedman, W. L., & Madore, B. F. 1990, ApJ 365, 186

Goldreich, P., & Tremaine, S. 1979 AJ 84, 1638

Goodman, J., & Lee, H. M. 1989, ApJ 337, 84

Gurzadyan, V. G., & Ozernoy, L. M. 1979, Nature 280, 214

Hut, P., & Rees, M. J. 1992, MNRAS 259, 27p

Kent, S. M. 1989, AJ 97, 1614

King, I. R., Stanford, S. A., & Crane, P. 1995, AJ 109, 164

Kormendy, J. 1988 ApJ 325, 128

Lacey, C. G., & Ostriker, J. P. 1985, ApJ 299, 633

Lauer, T. R., Faber, S. M., Groth, E. J., Shaya, E. J., Campbell, B., Code, A., Currie, D. G., Baum, W. A., Ewald, S. P., Hester, J. J., Holtzman, J. A., Kristian, J., Light, R. M., Lynds, C. R., O'Neil, E. J., & Westphal, J. A. 1993, AJ 106, 1436 (L93)

Lauer, T. R., Kormendy, J., Ajhar, E. A., Byun, Y.-I., Dressler, A., Faber, S. M., Grillmair, C., Richstone, D., & Tremaine, S. 1995, in preparation

Light, E. S., Danielson, R. E., & Schwarzschild, M. 1974, ApJ 194, 257

van der Marel, R. P. 1994, ApJ 432, L91

van der Marel, R. P., Rix, H.-W., Carter, D., Franx, M., White, S.D.M., & de Zeeuw, T. 1994, MNRAS 268, 521

Moore, B. 1993, ApJ 413, L93

Mould, J., Graham, J., Matthews, K. Soifer, T., & Phinney, E. S. 1989, ApJ 339, L21

Richstone, D. O., Bower, G., & Dressler, A. 1990, ApJ 353, 118

Stark, A. A. 1977, ApJ 213, 368

Syer, D., & Clarke, C. J. 1992, MNRAS 255, 92

Tremaine, S., & Weinberg, M. D. 1984, MNRAS 209, 729

Xu, G., & Ostriker, J. P. 1994, ApJ 437, 184